\documentclass[prb,twocolumn,showpacs]{revtex4}

\usepackage{graphicx}
\usepackage{dcolumn}
\usepackage{amsmath}

\begin{document}

\title[Neutron and X-ray synchrotron scattering]
  {Inelastic Neutron and X-ray Scattering from Incommensurate Magnetic Systems}

\author{Peter B\"oni$^1$, Bertrand Roessli$^2$, Klaudia Hradil$^3$}

 \affiliation{$^1$ Physik Department E21, Technische Universit\"at M\"unchen,
          D-85748 Garching, Germany}
 \affiliation{$^2$ Laboratory for Neutron Scattering, Paul Scherrer Institut,
          CH-5232 Villigen PSI, Switzerland}
 \affiliation{$^3$ X-Ray Center, Vienna University of Technology
          A-1060 Vienna, Austria}

\begin{abstract}
Neutrons and X-rays are powerful probes for studying magnetic and
lattice excitations in strongly correlated materials over very
wide ranges of momentum and energy transfers. In the focus of the
present work are the incommensurate magnetic systems MnSi and Cr.
Under application of a magnetic field, helically ordered MnSi
transforms into a weak itinerant ferromagnet. Using polarized
neutrons we demonstrate that the Stoner excitations are spin flip
excitations. The amplitude (longitudinal) fluctuations associated
with the magnon modes are already strong far away from $T_C$.
Interestingly, even the non spin flip excitations associated with
the Stoner modes are observable. In Cr, we have observed Kohn
anomalies in the phonon spectrum at those incommensurate positions
in reciprocal space, where the spin density wave is observed. The
corresponding phonon and magnon modes are not coupled. In
addition, an anomalous softening of a transverse phonon branch
along the N-H zone boundary line is observed that is caused by
strong electron phonon coupling. High resolution neutron
scattering indicate that the low energy Fincher-Burke excitations
may rather correspond to localized modes in momentum and energy
and not to propagating collective modes. Finally, we demonstrate
that in the near future it may become feasible to investigate
excitations in very small samples thus allowing to measure the
dynamics of strongly correlated materials under extreme conditions
and in the vicinity of quantum phase transitions.
\end{abstract}

\maketitle

\section{Introduction}

Effects related to incommensurate magnetic or charge order have
revealed many interesting effects in condensed matter physics.
Recent examples include i) high-$T_c$ superconductors, where
antiferromagnetic fluctuations may be responsible for the pairing
of the electrons \cite{tranquada2004}, ii) multiferroic compounds
such as the manganites RMnO$_3$ (R: lanthanide, alkaline metals)
\cite{tokura2003} or borates \cite{janoschek2010a}, where a
coupling between magnetic spiral-like order and the lattice or
magneto-elastic coupling may lead to ferro-electric coupling, or
iii) itinerant magnets such as MnSi, where a skyrmion lattice has
recently been identified \cite{muehlbauer2009}. By measuring the
collective excitations in these materials it is possible to
determine the energy scales that are responsible for the competing
interactions, which are the origin for the novel incommensurate
orderings.

Two prototypical magnetic systems are in the focus of our present
interest, namely MnSi and Cr. MnSi serves as a prototype system
for a weak itinerant magnet exhibiting modulated magnetically
ordered phases \cite{muehlbauer2009} due to the competition
between the Dzyaloshinskii Moriya and the exchange interaction
\cite{dzyal1958,moriya1960}. The incommensurate antiferromagnet Cr
is brought into focus due to the fact that the excitation spectrum
shows striking similarities with the magnetic excitations in high
$T_c$-superconductors \cite{endoh2006}. In order to access the
important energy($E$)-scales in these systems, various scattering
techniques are to be used.

With the ongoing efforts to improve the $E$-resolution of
inelastic X-ray beamlines approaching 1 meV, photons have become a
valuable tool for investigating the lattice dynamics in solid
state physics. In contrast to neutron scattering, the relative
change of the energy of the photons during scattering is small,
i.e. the scattering triangle remains essentially isosceles.
Therefore, the phonon dispersions can be measured efficiently over
large regions in momentum($\bf Q$)-space using area detectors.
Compared to investigations with photons, the strengths of neutron
scattering are a widely tunable $E$-resolution and the large
interaction with the magnetic degrees of freedom. To take benefit
of the individual probes, we have applied both techniques to
investigate the dynamics in Cr and MnSi as described below.

\section{Magnetic scattering from itinerant ferromagnets}

\begin{figure}[htb]   
 \includegraphics[scale = 0.4]{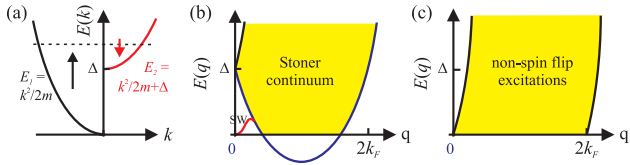}
 \centering
 \caption{The spin split bands are separated by the exchange splitting $\Delta$
  as shown in (a).
  (b) shows the continuum of spin flip excitations between
  the two bands, i.e. the Stoner continuum. The spin wave (SW) branch extends from
  $q = 0$ towards the Stoner continuum. (c) shows the continuum
  of non spin flip excitations within a band. }
 \label{MnSi_bands}
\end{figure}

First we describe inelastic neutron scattering experiments that
were performed with polarized neutrons in MnSi, thus providing a
direct means to observe single particle excitations in itinerant
ferromagnets. According to the most simple model for
ferromagnetism in delocalized systems, Stoner \cite{stoner1938}
assumed that within a single band model the interaction between
the spin up and spin down electrons leads to a separation of the
bands by an exchange splitting $\Delta$ (Fig.
\ref{MnSi_bands}(a)). The energy gain is partly compensated by an
increase of the kinetic energy of the conduction electrons. In
this picture, long range ferromagnetic order is destroyed by the
thermal excitation of electrons between the spin split bands. Fig.
\ref{MnSi_bands}(b) depicts the continuum of single particle
excitations between the spin up and spin down bands and (c) shows
the continuum of excitations within the bands. Also indicated in
(b) is the spin wave branch of the collective excitations
(magnons) that proceeds from $q = 0$ (Goldstone mode) towards the
Stoner continuum.

Extensive measurements of the spin fluctuations in the ordered
phase of MnSi using unpolarized neutrons have been performed by
Ishikawa et al. many years ago \cite{ishikawa1977}. These
experiments were not sensitive to distinguish between spin flip
and non spin flip contributions. In order to separate the
contributions, we have investigated the magnetic excitation
spectrum using inelastic neutron scattering with longitudinal
polarization analysis.

\begin{figure}[htb]   
 \centering
 \includegraphics[scale = 0.4]{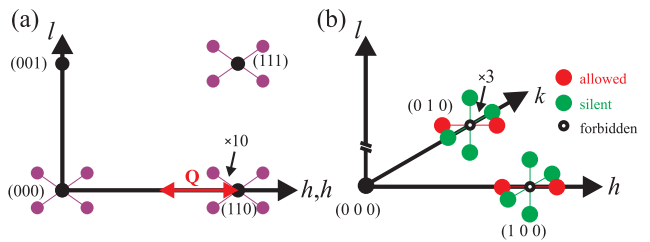}
 \caption{(a) Reciprocal lattice of helimagnetic MnSi showing the
 position of the magnetic satellites close to the nuclear Bragg
 reflections. (b) depicts the reciprocal lattice of
 Cr when it is prepared in a single ${\bf Q}$ state with $\bf Q$
 parallel to the [1 0 0]-direction. Therefore,
 magnetic satellites that are allowed by symmetry along
 the [0 1 0] and [0 0 1] directions are silent. The
 incommensurabilities are exaggerated by factors of 10 and 3 for MnSi
 and Cr, respectively.}
 \label{MnSi_cr_rec_space}
\end{figure}

Due to the lack of a symmetry center in the cubic crystal
structure ($a_{MnSi} = 4.56$ \AA) of MnSi (P2$_1$3), a left-handed
magnetic spiral with a long period $\Lambda \simeq 185$ \AA\ is
observed leading to magnetic satellite peaks close to the nuclear
reflections (Fig. \ref{MnSi_cr_rec_space}) \cite{tanaka1985}. Weak
crystal electric fields pin the spirals along the $\langle 1\, 1\,
1\rangle$ directions. Under application of a field $B \simeq 0.6$
T a ferromagnetic state is induced. Close to $T_C$ and in a field
$B \simeq 0.2$ T, a skyrmion lattice develops that is stabilized
by low energy fluctuations \cite{muehlbauer2009}. Apart from these
modulated phases, the magnetic properties are considered to be
those of a ferromagnet.

\begin{figure}[htb]
 \centering
 \includegraphics[scale = 0.8]{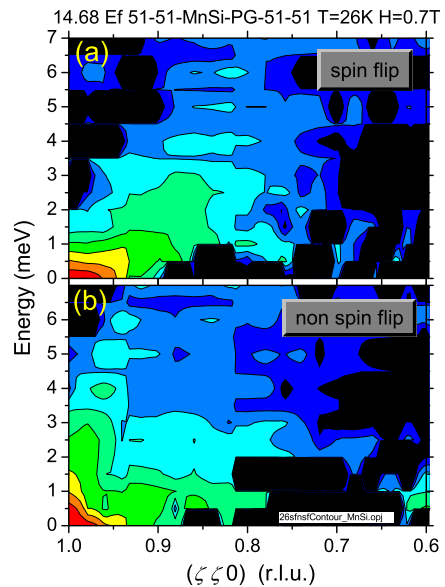}    
 \caption{Contour maps of MnSi measured along the [1 1 0] direction at $T = 26$ K and in a field
  $B = 0.7$ T.
  The sf data (a) clearly indicates the spin wave branch merging into the Stoner
  continuum. The contours of the nsf scattering (b) show
  the steep longitudinal phonon branch as well as the longitudinal fluctuations, which extend to high energy transfers.}
 \label{MnSi_sf_nsf_Contour}
\end{figure}

Fig. \ref{MnSi_sf_nsf_Contour} shows contour maps of the spin flip
(sf) and non spin flip (nsf) excitations of MnSi as measured in
the field induced ferromagnetic state ($B = 0.7$ T) at $T = 26 K$
($0.88\, T_C$). The experiments were performed on the triple axis
spectrometer IN20 at the Institut Laue-Langevin (ILL) around the
(1 1 0) Bragg peak in a (1 1 0) plane using longitudinal
polarization analysis (Fig. \ref{MnSi_cr_rec_space}).

The sf-data in Fig. \ref{MnSi_sf_nsf_Contour}(a) clearly shows the
spin wave branch emerging from the (1 1 0) Bragg peak becoming
very steep near $\zeta = 0.8$. One can clearly distinguish two
different regimes: i) at low $E$-transfer $E < 2.5$ meV ($0.9 <
\zeta < 1.0$) a ferromagnetic spin wave dispersion is observed
given by $E_q = Dq^2$ with $D = 23.5\pm3.0$ meV\AA$^2$
\cite{tixier1998} and ii) at large $E$-transfers the excitations
can be directly identified in terms of Stoner excitations as
depicted in Fig. \ref{MnSi_bands}(b). A close inspection of the
data shows that only the spin waves renormalize with increasing
$T$, while the Stoner excitations do not change significantly
\cite{semadeni1999}.

The nsf-data (Fig. \ref{MnSi_sf_nsf_Contour}(b)) shows at low
energy transfers large cross sections. They are identified as
longitudinal magnetic fluctuations, which have also been observed
in Ni \cite{boeni1991} and EuS \cite{boeni2002}. However, due to
the strong electronic correlations in MnSi leading to a large
magnetic correlation length, the longitudinal modes are already
strong much further away from $T_C$ namely already at $0.88\,
T_C$, where the present measurements have been conducted. These
modes diverge near $T_C$ due to the increasing magnon-magnon
interactions \cite{villain1970}. The longitudinal modes extend
into the regime of single particle excitations. The distribution
of their spectral weight looks qualitatively similar as the simple
model shown in Fig. \ref{MnSi_bands}(c).

\begin{figure}[htb]      
 \centering
 \includegraphics[scale = 0.6]{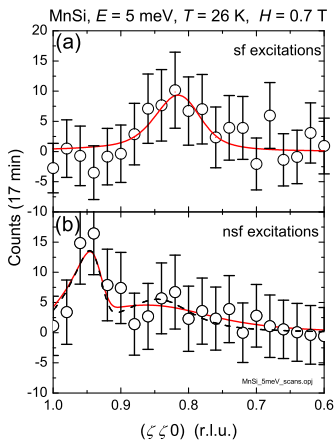}
 \caption{Constant energy scans for $E = 5$ meV at $T = 26$ K and $B = 0.7$ T. (a) shows
  a Stoner excitation near (0.8, 0.8, 0). The solid line
  is a fit to equation (\ref{doubleLorentzian}). (b) The magnetic nsf scattering is barely visible.
  The solid and the broken lines are fits to equation (\ref{doubleLorentzian})
  assuming a diffusive or a propagating mode, respectively, and a
  cross section for the acoustic phonon at $\zeta = 0.95$.}
 \label{MnSi_5meV_scans}
\end{figure}

To obtain a more quantitative interpretation of the data we show
in Fig. \ref{MnSi_5meV_scans} cuts through the contours of Fig.
\ref{MnSi_sf_nsf_Contour} at $E = 5$ meV. The Stoner excitation in
the spin flip channel is clearly visible near $\zeta = 0.83$. The
nsf scattering is very weak. The data sets were fitted using a
double Lorentzian scattering function

\begin{equation}
 S(q,E) = {q_d^2\over \kappa_1^2+q^2}{1\over 2\pi}
  {E\Gamma\over (E-E_q)^2 + \Gamma^2} \langle n+1 \rangle
 \label{doubleLorentzian}
\end{equation}
convoluted with the resolution function, where $\Gamma$ is the
linewidth parameterized by $\Gamma = Aq^{2.5}$ and $\kappa_1$ is
the inverse correlation length \cite{ishikawa1985}. The Lorentzian
is centered at the spin wave energy $E_q = Dq^2$, where $D$ is the
stiffness. In the case of quasielastic scattering, $D = 0$ and
$\kappa_1$ assumes a finite value \cite{boeni2002}. $\langle n+1
\rangle$ is the thermal population factor. The fit of the sf-data
demonstrates that the parameter $D_{sf} = 35$ meV\AA$^2$ for the
Stoner excitations is indeed significantly larger than the
stiffness parameter of the spin wave mode at low $E$-transfers,
$D_{sw} = 23.5$ meV \cite{tixier1998}. In contrast, the best fits
to the nsf-data converge towards the parameters $D_{nsf} = 0$ and
$A_{nsf} = 44$ meV\AA$^{2.5}$. $A_{nsf}$ is about twice as large
as the parameter $A = 19.6$ meV\AA$^{2.5}$ measured in the
paramagnetic phase of MnSi at small momentum and $E$-transfer
\cite{ishikawa1985}. This large $A_{nsf}$ is in line with the
increased stiffness of the Stoner excitations. A forced fit to the
data fixing $D_{nsf}$ at 23.5 meV leads to an increased $\chi^2$
(broken line in Fig. \ref{MnSi_5meV_scans}). It does not represent
the data well.

\begin{figure}[htb]
 \centering
 \includegraphics[scale = 0.3]{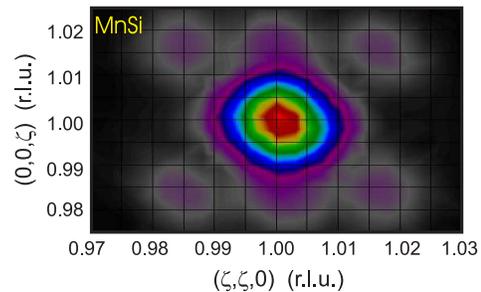}
 \caption{Elastic neutron scattering around the (1 1 0) Bragg peak
   showing the magnetic
  satellite peaks along the $\langle 1 1 1 \rangle$ directions. The weak spots at
  (1, 1, 0.985) and (1, 1, 1.015) are the result of the out of plane satellites, which
  become visible due to the coarse vertical resolution.}
 \label{MnSi_contour_Nov_22_02}
\end{figure}

We point out that the measurements shown here have been conducted
at very large $|{\bf q}| \gg 2\pi/\Lambda$, where the helical
correlations are expected to be of no relevance. Indeed, the
contour plot for $E=0$ (Fig. \ref{MnSi_contour_Nov_22_02})
demonstrates that the magnetic satellites are very close to the
nuclear zone centre.

Driven by the discovery of a non-Fermi liquid state under high
pressure and a skyrmion lattice for $B = 0.2$ T, the magnon
spectrum in the helical phase has recently been investigated with
high resolution triple axis spectroscopy using cold neutrons.
Around the magnetic satellites, a rich spectrum of spin
excitations are observed that are identified as helimagnons
\cite{janoschek2010b}. Using a model \cite{belitz2006} based on
only three parameters, namely the pitch of the helix, the spin
wave stiffness, and an overall amplitude of the signal one can
account for all spectra, demonstrating that helimagnons are a
universal characteristics of systems with weak chiral
interactions.

In conclusion, we have shown that inelastic neutron scattering
with polarized neutrons allows to study the single particle
excitations in weak itinerant magnets. The results prove directly
that the Stoner excitations are spin flip excitations.
Unexpectedly, the nsf fluctuations show significant cross sections
even at high energy transfers. To disentangle the complicated
spectra and to determine the helicity of the helimagnons it would
be of great value to extend the polarized beam measurements to
small momentum transfers using high resolution spectroscopy.
Indeed, recently it was shown that the paramagnetic
\cite{roessli2002} and helimagnetic \cite{roessli2004} excitations
in MnSi have a chiral contribution, which is large even close to
$T_C$.

\section{Magnon and Phonon Excitations in Antiferromagnetic Cr}

In metals with Fermi surfaces, nesting enhances the number of
transitions at the nesting wavevectors ${\bf Q}_n$ when compared
to other wavevectors. The nesting greatly increases the number of
possible electronic transitions at ${\bf Q}_n$, which may lead to
the formation of spin density waves (SDW)
\cite{overhauser1960,overhauser1962} and/or may soften and broaden
phonons \cite{kohn1959}. Both effects are observed in strongly
correlated electron systems, i.e. in the copper oxide
superconductors \cite{djajaputra1999,reznik2010} and in elemental
Cr \cite{fawcett1988}. Both materials show a long history of
intensive investigations.

\begin{figure}
 \centering
 \includegraphics[scale = 0.3]{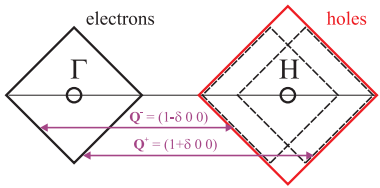}   
 \caption{Schematic cross section of the Fermi surface in the (1 1 0)
 plane of Cr. Nesting vectors $\bf Q$ connect the electron with
 the hole surface.}
 \label{Cr_Fermi_Surface}
\end{figure}

In contrast to MnSi, where the magnetic ordering is caused by the
non-centrosymmetric crystal structure leading to a pronounced
Dzyaloshinskii-Moriya (DM) interaction
\cite{dzyal1958,moriya1960}, the incommensurate magnetic ordering
in Cr is the result of the nesting properties of the electron and
hole Fermi surface \cite{overhauser1962} as explained in Fig.
\ref{Cr_Fermi_Surface}. Despite numerous investigations, the
magnetic excitations in Cr are still not well understood and some
of the controversial experimental results concerning the
Fincher-Burke (FB) modes \cite{burke1983} have only recently been
reinvestigated \cite{hiraka2003}.

Cr crystallizes in a simple body centered cubic structure with a
lattice constant $a_{Cr} = 2.88$ \AA. It  undergoes a transition
from the paramagnetic phase to a SDW phase at $T_N = 311$~K
characterized by propagation vectors ${\bf Q}^\pm =(1 \pm
\delta,\,0,\, 0)$ with $\delta = 0.048$. The corresponding
magnetic satellite peaks are visible near the forbidden Bragg
reflections of the bcc structure, i.e. $h+k+l = \rm odd$ (Fig.
\ref{MnSi_cr_rec_space}). In the transverse spin-density wave
(TSDW) phase $T_{sf} < T < T_N$ the magnetic moments are aligned
perpendicular to $\mathbf{Q}^\pm$. At $T_{sf} = 121~{\rm K}$ the
magnetic structure undergoes a first order phase transition to a
longitudinal spin-density wave (LSDW) phase with the spins aligned
along $\mathbf{Q}^\pm$.

Due to the SDW, there is a distortion of the lattice with twice
$\delta$. Indeed, charge density waves (CDW) were observed using
both neutron and X-ray diffraction\cite{pynn1976,tsunoda1974}.
Recently even pressure measurements were performed
\cite{jaramillo2009}. The CDW can be induced either by Fermi
surface nesting or by a strain wave induced by magnetoelastic
coupling to the SDW.

\begin{figure}
 \centering
 \includegraphics[scale = 1.0]{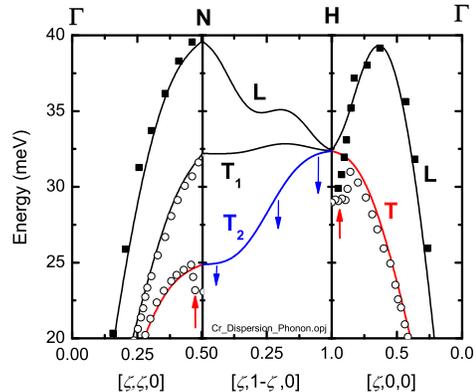}
 \caption{Comparison of the phonon dispersion as calculated by means of
 the Born-von-K\'arm\'an model along the high symmetry directions $[\zeta\, \zeta\, 0]$
 and $[\zeta\, 0\, 0]$. The data points from reference \cite{shaw1971} show strong
 anomalies near the N and H points as indicated by red arrows. Our results show that the T2 branch
 along the whole zone boundary line N-H softens as indicated by three blue arrows.}
 \label{Cr_Dispersion_Phonon}
\end{figure}

As explained above, the interaction of the conduction electrons
with the lattice vibrations enhanced by anomalies of the Fermi
surface leads to anomalous phonon dispersions. Using inelastic
neutron scattering, four regions have been identified where
transverse acoustic phonons show anomalous behavior that can be
traced back to nesting \cite{shaw1971}. The two most pronounced
anomalies occur near the N- and H-point as shown in Fig.
\ref{Cr_Dispersion_Phonon}. Because of the appearance of a SDW
near H, this point is of particular interest. However, because of
the coarse $\bf Q$-resolution of neutron scattering the phonon
anomaly may be washed out. Here, the improved $\bf Q$-resolution
of synchrotron radiation may help to highlight if there is a
correspondence between the phonon anomalies and Fermi surface
nesting in Cr near H.

\subsection{Magnetic excitations in Cr}

\begin{figure}
 \centering
 \includegraphics[scale = 0.35]{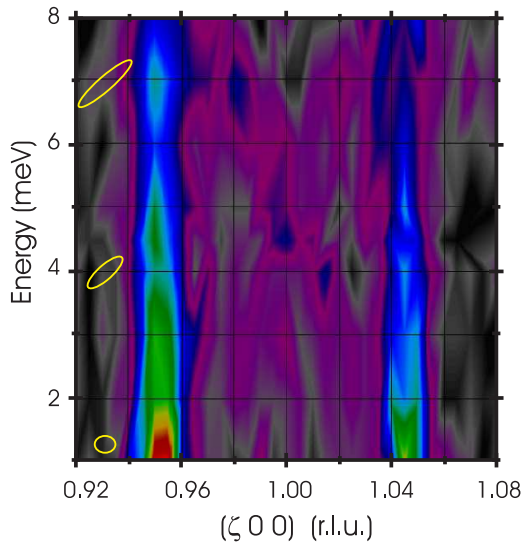}
 \caption{Contour map of the excitation spectrum as measured in the
 transverse spin density wave phase of Cr at $T = 136$ K. The
 intense peak near 4.5 meV and ${\bf Q} = (1,\,0,\,0)$ is the Fincher mode.
 Note the additional peak near $(1.02,\,0,\,0)$ and 3 meV that has
 no counter part at $(0.98,\,0,\,0)$. The ellipsoids indicate the
 resolution of the triple axis spectrometer TASP
 \cite{semadeni2001} for various $E$-transfers.}
 \label{Cr_136K_contour}
\end{figure}

In order to obtain an overview on the spectral distribution of the
magnetic excitations in Cr we show in Fig.~\ref{Cr_136K_contour} a
contour plot of the inelastic intensity in the TSDW-phase at 136 K
that was measured with high energy and momentum resolution using
cold neutrons ($E_f = 5.64$ meV). Most dominant are the very steep
excitations that emerge from the incommensurate positions. Because
of the very large velocity of the excitations, $c \simeq 1500$
meV\AA, the dispersion branches cannot be resolved
\cite{fincher1981,boeni1998}. Taking into account that various
simple band models predict a velocity $c \simeq v_F/\sqrt{3}$
where $v_F$ is the Fermi velocity \cite{fedders1966}, this high
$c$ value seems reasonable. At high energy transfers, $E > 30$ meV
the dispersion branches bend towards the commensurate position
\cite{fincher1981} because of the increasing spectral weight of
the phason modes with increasing energy
\cite{fishman1996,endoh2006}.

\begin{figure}[htb]
 \centering
 \includegraphics[scale = 0.7]{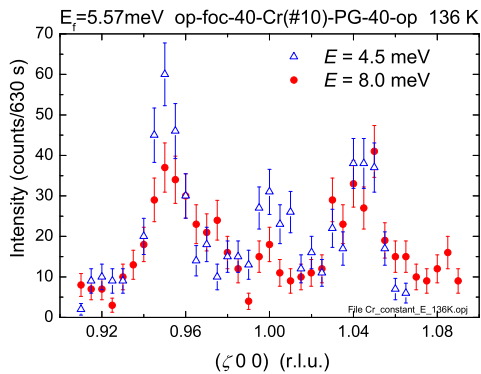}
 \caption{Constant-$E$ scans showing the Fincher mode at (1, 0, 0) for $E
 = 4.5$ meV and a weak mode at 8 meV that was already observed in
 Ref. \cite{fincher1981}.}
 \label{Cr_Escan_4_8}
\end{figure}

Other important features are the low-energy excitations at $E <
10$ meV, which occur only in the TSDW phase
\cite{hiraka2003,hiraka2004}. Clearly visible are the modes at $E
\simeq 4.5$ meV and at 8 meV (Fig. \ref{Cr_Escan_4_8}) as already
observed by Fincher et al. demonstrating the correspondence of our
data with the previous data \cite{fincher1981}. In addition, two
modes at $\simeq 1.016$ \AA$^{-1}$/$\simeq 3.8$ meV and $\simeq
0.984$ \AA$^{-1}$/$\simeq 6.8$ meV are visible, which have no
counter parts at the symmetry related positions $\simeq 0.984$
\AA$^{-1}$ and $\simeq 1.016$ \AA$^{-1}$, respectively. The
comparison with previous data measured with lower resolution
\cite{burke1983,hiraka2004} raises the question if dispersing
modes alone can explain the data or if local modes have to be
included.

Moreover, it is not easily understandable why the measured modes
have very similar intensities although the thermal population
factor $\langle n +1\rangle$ would predict a significantly higher
intensity for the 3.8 meV mode when compared with the 6.8 meV mode,
provided that both modes belong to the same dispersion as alluded to in Ref.
\cite{burke1983}. We may speculate that mode-coupling has to be involved. To
answer these questions, more precise measurements with better
statistics have to be performed.

\subsection{Phonon softening in Cr}
\label{phonons}

The goal of the experiments using synchrotron radiation was to
determine the precise $\bf q$-position of the anomalous softening
of the phonons and to compare it with the nesting features of the
Fermi surface of Cr. In order to improve in $\bf Q$-resolution
over the previous experiments using inelastic neutron scattering
\cite{shaw1971}, we applied inelastic X-ray scattering. The
experiment was performed at beam line ID-28 at ESRF using a Cr
single crystal with the dimensions $2\times2\times2$ mm$^3$. The
phonon dispersion was investigated near the H-point and the
measurements extended along the zone boundary to the N-point
\cite{lamago2010}. In addition to previous measurements, the
phonon dispersion along the line connecting the H-point and the
N-point (Fig. \ref{Cr_Dispersion_Phonon}) was explored for the
first time.

\begin{figure}
 \centering
 \includegraphics[scale = 0.5]{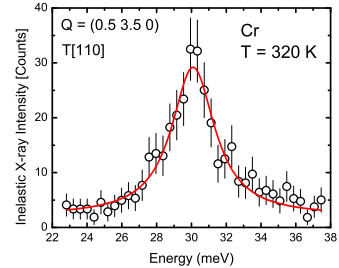}  
 \caption{Raw data from inelastic X-ray scattering as function of
  energy transfer at $T = 320$ K and ${\bf Q} = (0.5, 3.5, 0)$.
  The solid line corresponds to a fit assuming a Lorentzian.}
 \label{pboeni_Cr_Phonon_ESRF}
\end{figure}

Fig. \ref{pboeni_Cr_Phonon_ESRF} shows a phonon at ${\bf Q} =
(0.5, 3.5, 0)$ that belongs to the acoustic [1 1 0] T2 branch.
Obviously, the position of the phonons can be measured with high
precision. The solid line is a fit assuming a Lorentzian.
Following this result, the dispersion of the phonons was
determined in detail for $\bf Q$ along [1 0 0] as well as along
the N-H line \cite{lamago2010}.

\begin{figure}
 \centering
 \includegraphics[scale = 0.7]{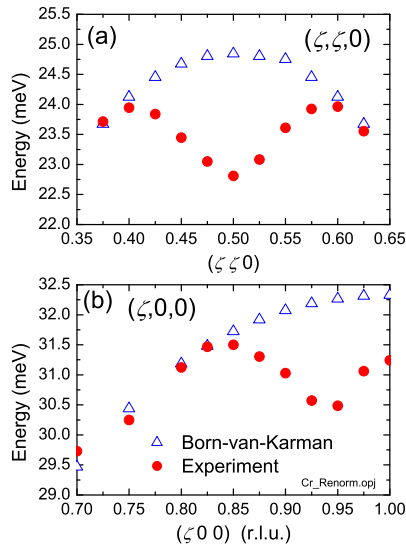}
 \caption{(a) and (b) show the dispersion of the phonons near the N and H-points,
 respectively. The phonons along $[\zeta\,\zeta\, 0]$ soften directly at the N-point
 while the phonons along $[\zeta\,0\,0]$ soften at the nesting
 position, where the SDW ordering is observed (Fig. \ref{Cr_136K_contour}).}
 \label{Cr_Renorm}
\end{figure}

The measured dispersions near $({1\over 2}\, {1\over 2}\, 0)$ and
(1 0 0) are shown in Fig. \ref{Cr_Renorm} (circles) and compared
with the theoretical prediction based on a simple Born-von-Karman
model (triangles). Along the [1 1 0] direction, the maximum
softening is observed exactly at the zone boundary N. In contrast,
the H-phonon softens at the incommensurate position $\delta =
0.05$ where the SDW satellites occur and not at the zone boundary
(1 0 0) as reported previously (Fig. \ref{Cr_Dispersion_Phonon})
\cite{shaw1971}.

\begin{figure}
 \centering
 \includegraphics[scale = 0.6]{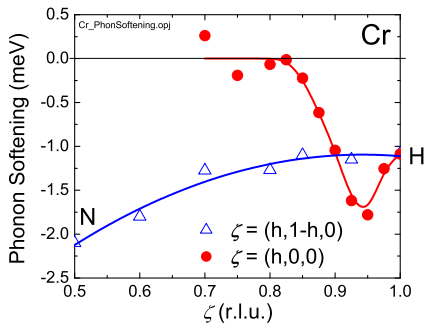}
 \caption{Difference $E_{BvK} - E_{exp}$ between the
 Born-von-K\'arm\'an
 and the experimental phonon dispersions along the zone boundary N-H
 (triangles) and along the high symmetry direction $\Gamma$-H (circles).}
 \label{PhonSoftening}
\end{figure}

Similar measurements have been conducted at various $\bf
Q$-positions across the zone boundary line N-H. Fig.
\ref{PhonSoftening} summarizes the softening of the measured
phonons when compared with a Born-von-K\'arm\'an model along the
[1 0 0] direction (circles) and along the zone boundary N-H
(triangles). In the [1 0 0] direction, the phonon softening has a
distinct minimum at the nesting wavevector ${\bf Q^\pm} = (0.95,
0, 0)$. Surprisingly, a strong anomaly also appears along the
entire zone boundary line N-H indicating that strong electron
phonon coupling limited to a small range of wavevectors can also
result in strong phonon anomalies without invoking nesting. This
observation implies that the phonon anomalies in copper oxide
superconductors may also be explained by an enhanced
electron-phonon coupling without invoking novel collective modes
or some hidden nesting of the Fermi surface \cite{lamago2010}.

Additional measurements in the paramagnetic phase of Cr show a
similar softening of the phonons close to the zone boundaries
discussed above. In particular, the strong softening at ${\bf
Q^\pm} = (0.95, 0, 0)$ persists demonstrating that the
magneto-elastic coupling is small. Therefore, the nesting of the
Fermi surface is responsible for the Kohn anomaly in the phonon
spectrum and the SDW below $T_N = 311$ K at $\bf Q^\pm$. They
evolve independently with temperature.

In conclusion, we have shown that the incommensurate magnetic
excitations in Cr can be interpreted in terms of electron hole
excitations at the Fermi surface. Strong electron-phonon coupling
without invoking nesting leads to a pronounced softening of the T2
phonon branch along N-H, i.e. away from $\bf Q^\pm$. We have
provided evidence that the low energy excitations in the TSDW
phase may not be explained in terms of a mode having a dispersion
as proposed in Ref. \cite{burke1983}.

\section{Small Samples - Extreme Conditions}

Studying quantum phase transitions by applying pressure, magnetic
fields or by doping has proven to be a successful route to
identify materials with novel properties. In MnSi, the application
of pressure leads to the suppression of helical order near $p_c =
1.46$ GPa accompanied by the appearance of a partially ordered
magnetic state and a non Fermi liquid phase \cite{pfleiderer2004}.
In Cr, antiferromagnetic order is suppressed around $p = 10$ GPa
\cite{jaramillo2009}. Neutron scattering as well as X-ray
synchrotron scattering have provided valuable information about
the vanishing of the order parameters, however, the present day
sensitivity of neutron spectrometers is not sufficient to also
characterize the spectrum of the magnetic and lattice excitations
close to the quantum phase transitions because the samples are
usually very small \cite{niklowitz2009}.

Measurements of phonon dispersions may be conducted using
inelastic X-ray scattering because only small samples are
necessary and the $\bf Q$- and $E$-resolutions are often
sufficient as shown in section \ref{phonons}. Inelastic magnetic
scattering by X-rays is so far impossible because the magnetic
cross sections are very small. In the following we demonstrate the
feasibility of inelastic neutron scattering on small samples using
supermirror focusing guides \cite{boeni2008}, to generate very
small but intense neutron beams at the sample position of the
thermal triple axis instrument PUMA at FRM II.

To perform the experiment, a focusing guide with a length of 500
mm as described in Ref. \cite{muehlbauer2006} has been installed
between the monochromator and the sample position. In the first
instance we used a neutron CCD camera to both correctly align the
guide and to analyze the shape and intensity of the beam at the
sample position.  These measurements show that the beam has a FWHM
of approximately 2 mm in the horizontal direction and 8 mm in the
vertical direction at the sample position. Such a beam size is
ideal for studying samples with mm dimensions, unlike the
conventional PUMA profile where the primary beam has dimensions of
roughly 25 mm in the horizontal direction and 28 mm in the
vertical direction and any adjustment for sample size is done by
adjustable, neutron absorbing slits. The CCD camera images also
revealed that using the guide results in a very low background.

\begin{figure}[htb]
 \centering
 \includegraphics[scale = 0.6]{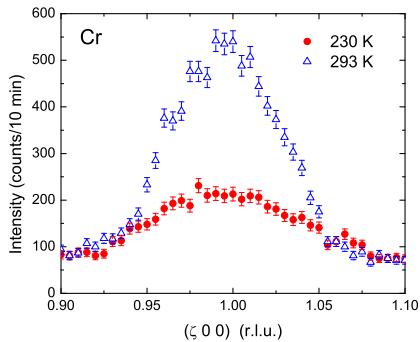}
 \caption{Temperature dependence of the magnetic excitations in Cr as measured at
 $T = 230$ K and 293 K around the (1 0 0) Bragg reflection at an
 energy of 4 meV.}
 \label{Cr_Focusing}
 \end{figure}

We have performed successful test experiments on two different
samples.  Firstly, a small single crystal of Cr with dimensions
$2\times 2\times 2$ mm$^3$ was investigated. Samples of this size
can typically be used in a Paris-Edinburgh high pressure cell.
Using the focusing guide, we were able to observe amongst other
things the change in intensity of the magnetic excitations in the
TSDW phase with increasing temperature (Fig.~\ref{Cr_Focusing}).
Note that the counting time per point of 10 min corresponds to the
typical counting time (13 min) used to collect the data in Fig. 3
of Ref. \cite{fincher1981}. Clearly, because of the coarse $\bf
Q$-resolution due to the focusing, the fine structure of the
spectrum is wiped out. Still, for the determination of the energy
scale of the excitations versus pressure, inelastic measurements
are feasible.

\begin{figure}[htb]    
 \centering
 \includegraphics[scale = 0.7]{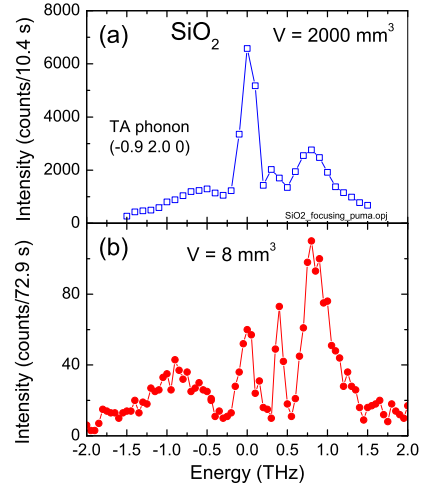}
 \caption{Measurements of a transverse acoustic phonon in quartz.  (a) Normal PUMA
 configuration using a doubly focusing monochromator. (b) Same measurement
 using a flat monochromator combined with a focusing guide. The
 small sample is 250 times smaller than the big sample.}
 \label{SiO2_Focusing}
\end{figure}

In a second experiment we measured phonons in two different
samples of quartz (SiO$_2$). The small and the large samples have
a volume of 8 mm$^3$ and 2000 mm$^3$, respectively
(Fig.~\ref{SiO2_Focusing}). The comparison allows a direct
calibration between the conventional PUMA configuration and the
configuration with a focusing guide. The results show that it is
indeed possible to measure the acoustic transverse phonon at
$(-0.9, 2.0, 0)$ with high precision within a reasonable time.
Actually the comparison of the two measurements demonstrates that
the results with the focusing guide provide a much cleaner
spectrum and a lower background. Of course, the performance can be
further increased without a significant loss in intensity if a
second focusing guide is used between the sample and the analyzer
\cite{niklowitz2009} and if the critical angle of reflection of
the focusing guide is increased from $m = 3$ to $m = 7$
\cite{snag}.

Concluding, we have shown that advanced neutron optics allows the
investigation of magnons and phonons under extreme conditions
using small crystals providing a good $E$- and $\bf Q$-resolution.
Focusing techniques for inelastic neutron scattering facilitate
the exploration of new areas of science, where traditional
experiments have been restricted due to the requirement of large
samples.

\section{Conclusions}

Experiments using polarized neutrons allow the separation of
phonon modes from magnon modes and furthermore between transverse
and longitudinal fluctuations in magnetic systems. In MnSi it is
observed that the Stoner excitations can be clearly identified in
terms of spin flip excitations of the conduction electrons. The
non spin flip scattering provides direct information on the
amplitude fluctuations in the single particle regime. In addition,
we found evidence that the low energy excitations in Cr
(Fincher-Burke modes) may not be explained solely by assuming
modes with a dispersion. Surprisingly the modes are asymmetric
with respect to the (1 0 0) position \cite{hiraka2004}. The
combination of inelastic X-ray and neutron scattering indicates
that the Kohn anomaly near the H-point occurs at the same position
as the spin density wave demonstrating that both effects are
caused independently by the nesting properties of the Fermi
surface of Cr. As a new feature, the X-ray results imply that the
phonons are softened along the whole zone boundary line N-H due to
electron-phonon interactions.

We have shown that the inelastic scattering of neutrons and X-rays
provide complementary information on the lattice dynamics and the
spectrum of the magnetic excitations in strongly correlated
materials . While neutron scattering provides high-energy
resolution and moderate $\bf Q$-resolution, X-rays provide
sufficient $\bf Q$-resolution thus allowing to map out the
softening of phonons near the zone boundaries of transition metals
like Cr.

Finally we have evaluated the possibilities of performing
inelastic neutron scattering from small samples using advanced
focusing techniques. By means of focusing guides the intensity of
neutron beams at the sample position can be increased rather
dramatically thus allowing the investigation of sample with a
volume of a few mm$^3$ \cite{hils2004}. Therefore, samples under
extreme conditions can be investigated near quantum phase
transitions. With the advancement of next generation neutron
sources using halo isomers that provide polarized neutron beams of
very high brilliance \cite{habs2011}, neutron scattering may make
a big leap towards the investigation of excitations in samples
that are much smaller than 1 mm$^3$.

\section*{ACKNOWLEDGMENTS}

We thank S. Dunsiger, M. Janoschek, D. Lamago, R. Mole, S.
M\"uhlbauer, P. Niklowitz, C. Pfleiderer, D. Reznik, and C.
Schanzer for very useful discussions and assistance during the
course of the experiments. This work is based on experiments
performed at the FRM II in Garching, the Swiss Spallation Source
SINQ, the HFR at the Institut Laue-Langevin, and the ESRF in
Grenoble. We gratefully acknowledge financial support from the
research unit on Quantum Phase Transitions FOR960 of the German
Science Foundation (DFG). Part of the work was supported by the
Swiss National Science Foundation through MaNEP and by the project
226507-NMI3 within the seventh framework program FP7 of the EU.



\begin{thebibliography}{0}%
\makeatletter
\providecommand \@ifxundefined [1]{%
 \@ifx{#1\undefined}
}%
\providecommand \@ifnum [1]{%
 \ifnum #1\expandafter \@firstoftwo
 \else \expandafter \@secondoftwo
 \fi
}%
\providecommand \@ifx [1]{%
 \ifx #1\expandafter \@firstoftwo
 \else \expandafter \@secondoftwo
 \fi
}%
\providecommand \natexlab [1]{#1}%
\providecommand \enquote  [1]{``#1''}%
\providecommand \bibnamefont  [1]{#1}%
\providecommand \bibfnamefont [1]{#1}%
\providecommand \citenamefont [1]{#1}%
\providecommand \href@noop [0]{\@secondoftwo}%
\providecommand \href [0]{\begingroup \@sanitize@url \@href}%
\providecommand \@href[1]{\@@startlink{#1}\@@href}%
\providecommand \@@href[1]{\endgroup#1\@@endlink}%
\providecommand \@sanitize@url [0]{\catcode `\\12\catcode `\$12\catcode
  `\&12\catcode `\#12\catcode `\^12\catcode `\_12\catcode `\%12\relax}%
\providecommand \@@startlink[1]{}%
\providecommand \@@endlink[0]{}%
\providecommand \url  [0]{\begingroup\@sanitize@url \@url }%
\providecommand \@url [1]{\endgroup\@href {#1}{\urlprefix }}%
\providecommand \urlprefix  [0]{URL }%
\providecommand \Eprint [0]{\href }%
\@ifxundefined \urlstyle {%
  \providecommand \doi  [0]{\begingroup \@sanitize@url \@doi}%
  \providecommand \@doi [1]{\endgroup \@@startlink {\doibase
  #1}doi:\discretionary {}{}{}#1\@@endlink }%
}{%
  \providecommand \doi  [0]{doi:\discretionary{}{}{}\begingroup
  \urlstyle{rm}\Url }%
}%
\providecommand \doibase [0]{http://dx.doi.org/}%
\providecommand \Doi [0]{\begingroup \@sanitize@url \@Doi }%
\providecommand \@Doi  [1]{\endgroup\@@startlink{\doibase#1}\@@Doi}%
\providecommand \@@Doi [1]{#1\@@endlink}%
\providecommand \selectlanguage [0]{\@gobble}%
\providecommand \bibinfo  [0]{\@secondoftwo}%
\providecommand \bibfield  [0]{\@secondoftwo}%
\providecommand \translation [1]{[#1]}%
\providecommand \BibitemOpen [0]{}%
\providecommand \bibitemStop [0]{}%
\providecommand \bibitemNoStop [0]{.\EOS\space}%
\providecommand \EOS [0]{\spacefactor3000\relax}%
\providecommand \BibitemShut  [1]{\csname bibitem#1\endcsname}%
\end{thebibliography}%


\begin{thebibliography}{10}
\bibitem{tranquada2004} Tranquada J et al. 2004 Nature {\bf 429} 534
\bibitem{tokura2003} Tokura Y 2003 Phys. Today {\bf 56} 50
\bibitem{janoschek2010a} Janoschek M, Fischer P, Schefer J, Roessli
    B, Pomjakushin V, Meven M, Petricek V, Petrakovskii G,
    Bezmaternikh L (2010) Phys. Rev. B {\bf 81} 094429
\bibitem{muehlbauer2009} M\"uhlbauer S, Binz B, Jonietz F, Pfleiderer C, Rosch A, Neubauer A, Georgii R, B\"oni P 2009 Science {\bf 323} 915
\bibitem{dzyal1958} Dzyaloshinskii L 1958 J. Phys. Chem. Solids {\bf 4} 241
\bibitem{moriya1960} Moriya T 1960 Phys. Rev. {\bf 120} 91
\bibitem{endoh2006} Endoh Y, B\"oni P 2006 J. Phys. Soc. Jpn. {\bf 75} 111002
\bibitem{stoner1938} Stoner E 1938 Proc. Roy. Soc. A {\bf 165} 372
\bibitem{ishikawa1977} Ishikawa Y, Shirane G, Tarvin J A, Kohgi M (1977) Phys. Rev. B {\bf 16} 4956
\bibitem{tanaka1985} Tanaka M, Takayoshi H, Ishida M, Endoh Y (1985) J. Phys. Soc. Jpn. {\bf 54} 3232
\bibitem{tixier1998} Tixier S, B\"oni P, Endoh Y, Roessli B, Shirane G 1999 Physica B {\bf 241-243} 613
\bibitem{semadeni1999} Semadeni F, B\"oni P, Endoh Y, Roessli B, Shirane G 1999 Physica B {\bf 267-268} 248
\bibitem{boeni1991} B\"oni P, Mart\'inez J L, Tranquada J M (1991) Phys. Rev. B {\bf 43} 575
\bibitem{boeni2002} B\"oni P, Roessli B, G\"orlitz D, K\"otzler J, Phys. Rev. B {\bf 65} 144434
\bibitem{villain1970} Villain J (1970) Solid State Commun. {\bf 8} 31
\bibitem{ishikawa1985} Ishikawa Y, Noda Y, Uemura Y J, Majkrzak C F, Shirane G (1985) {\bf 31} 5884
\bibitem{janoschek2010b} Janoschek M, Bernlochner F, Dunsiger S, Pfleiderer C, B\"oni P, Roessli B, Link P, Rosch A (2010) Phys. Rev. B {\bf 81} 214436
\bibitem{belitz2006} Belitz D, Kirkpatrick T R, Rosch A (2006) Phys. Rev. B {\bf 73} 054431
\bibitem{roessli2002} Roessli B, B\"oni P, Fischer W, Endoh Y 2002 Phys. Rev. Lett. {\bf 88} 237204
\bibitem{roessli2004} Roessli B, B\"oni P, Fischer W E, Endoh Y (2004) Physica B {\bf 345} 124
\bibitem{overhauser1960} Overhauser A W, Arrott A 1960 Phys. Rev. Lett.{\bf 4} 226
\bibitem{overhauser1962} Overhauser A W 1962 Phys. Rev. {\bf 128} 1437
\bibitem{kohn1959} Kohn W 1959 Phys. Rev. Lett. {\bf 2} 393
\bibitem{djajaputra1999} Djajaputra D and Ruvalds J 1999 Solid State Commun. {\bf 111} 199
\bibitem{fawcett1988} Fawcett E (1988) Rev. Mod. Phys. {\bf 60} 209
\bibitem{reznik2010} Reznik D 2010 Advances in Condensed Matter Physics {\bf 2010} 523549
\bibitem{burke1983} Burke S K, Stirling W G, Ziebeck K R A, Booth, J G 1983 Phys. Rev. Lett. {\bf 51} 494
\bibitem{hiraka2003} Hiraka H, Endoh Y, B\"oni P, Fujita M, Yamada K, Shirane G 2003 Phys. Rev. B {\bf 67} 064423
\bibitem{hiraka2004} Hiraka H, B\"oni P, Yamada K, Park S, Lee S-H, Shirane G 2004 Phys. Rev. B {\bf 70} 144413
\bibitem{pynn1976} Pynn R, Press W, Shapiro M S, Werner S A 1976 Phys. Rev. B {\bf 13} 295
\bibitem{tsunoda1974} Tsunoda Y, Mori M, Kunitomi N, Teraoka Y, Kanamori J 1974 Solid. State Commun. {\bf 14} 287
\bibitem{jaramillo2009} Jaramillo R, Feng Y, Lang J C, Islam Z, Srajer G, Littlewood P B, Whan D B, Rosenbaum T F 2009 Nature {\bf 459} 405
\bibitem{shaw1971} Shaw W M, Muhlestein L D 1971 Phys. Rev. B {\bf 4} 969
\bibitem{fincher1981} Fincher C R, Shirane G, Werner S A 1981 Phys. Rev. B {\bf 24} 1312
\bibitem{boeni1998} B\"oni P, Sternlieb B, Shirane G, Roessli B, Lorenzo J E, Werner S A 1998 Phys. Rev. B {\bf 57} 1057
\bibitem{fedders1966} Fedders P A, Martin P C (1966) Phys. Rev. {\bf 143} 245
\bibitem{fishman1996} Fishman R S, Liu S H 1996 Phys. Rev. B {\bf 54} 7252
\bibitem{semadeni2001} Semadeni F, Roessli B, B\"oni P (2001) Physica B {\bf 297} 152
\bibitem{lamago2010} Lamago D, Hoesch M, Krisch M, Heid R, Bohnen K-P, B\"oni P, and Reznik D (2010) Phys. Rev. B {\bf 82} 195121
\bibitem{pfleiderer2004} Pfleiderer C, Reznik D, Pintschovius L, v. L\"ohneysen H, Garst M, Rosch A (2004) Nature {\bf 427} 227
\bibitem{niklowitz2009} Niklowitz P G, Pfleiderer C, M\"uhlbauer S, B\"oni P, Keller T, Link P, Wilson J A, Vojta M, Mydosh J A (2009) Physica B {\bf 404}, 2955
\bibitem{boeni2008} B\"oni P (2008) Nucl. Instrum. Methods A {\bf 586} 1-8
\bibitem{muehlbauer2006} M\"uhlbauer S, Stadlbauer M, B\"oni P, Schanzer C, Stahn J, Filges U (2006) Physica B {\bf 385-386} 1247 (2006)
\bibitem{snag} http://www.swissneutronics.ch/products/coatings.html
\bibitem{hils2004} Hils T, B\"oni P, Stahn J (2004) Physica B {\bf 350} 166
\bibitem{habs2011} Habs D, Gross M, Thirolf P G, B\"oni P (2011) Appl. Phys. B., DOI10.1007/s00340-010-4276-3 (accepted for publication)
\end{thebibliography}
\end{document}